\documentclass[preprint,12pt]{elsarticle}

\usepackage{amssymb}
\usepackage{amsmath}
\usepackage[numbers]{natbib}
\usepackage{tikz} 

\usetikzlibrary{arrows.meta} 
\usetikzlibrary{decorations.markings}  
\tikzset{midarrow/.style={postaction={decorate,decoration={markings,mark=at position 0.74 with {\arrow{Latex[length=3mm]}}}}}}

\cortext[cor1]{Corresponding author}

\begin{document}

\title{Green's Function Formalism for Impurity-Induced Resonances in Sub-barrier Proton-Nucleus Scattering}

\author[1]{Bahruz Suleymanli}

\ead{bahruz.suleymanli@gmail.com}
\affiliation[1]{organization={Physics Department, Yıldız Technical University},
                city={Istanbul},
                postcode={34220}, 
                state={Esenler},
                country={Türkiye}}

\author[1,2]{Kutsal Bozkurt}
\affiliation[2]{organization={Université Paris-Saclay, CNRS/IN2P3, IJCLab},
                city={Orsay},
                postcode={91405}, 
                country={France}}

\begin{abstract}
Motivated by recent experimental refinements of stellar reaction rates, we establish a non-perturbative Green's function formalism based on the exact solution of the Dyson equation for sub-barrier proton-nucleus resonant scattering. By utilizing bare Green's functions to map the quantum tunneling problem onto a scattering formalism, we demonstrate that the summation of infinite quantum paths recovers the exact tunneling coefficients, enabling an analytical solution of the Dyson equation where the strong nuclear force is modeled as a surface delta-shell impurity embedded within the Coulomb field.
Applying this framework to the astrophysically relevant $p + {}^{7}\text{Li}$, $p + {}^{14}\text{N}$, and $p + {}^{23}\text{Na}$ systems, we achieve precise agreement with experimental resonance energies while revealing a fundamental physical distinction in resonance formation. The heavier ${}^{23}\text{Na}$ system is identified as a saturated state, residing on a geometric plateau where the resonance energy becomes insensitive to the interaction strength; our calculated value of $2.11$~MeV aligns remarkably well with the experimental level of $2.08$~MeV. In contrast, the lighter ${}^7\text{Li}$ and ${}^{14}\text{N}$ systems emerge as threshold states in a weak-coupling window, where the resonance energy is highly sensitive to the potential parameters and is sustained near the continuum edge. In this regime, our model yields energies of $0.489$~MeV and $1.067$~MeV, closely reproducing the experimental benchmarks of $0.441$~MeV and $1.058$~MeV, respectively. We demonstrate that these threshold states are characterized by a significant enhancement of the resonant cross-section, driven by the inverse relationship between the tunneling width and the spectral density peak. Finally, we establish the domain of validity for this method via a systematic lifetime scan across the periodic table ($Z=2-50$), identifying a sharp transition at $Z = 18$ (Argon). This finding confirms that while the method is bounded by the onset of classical stability in heavier nuclei, it provides a rigorous and parameter-free theoretical baseline for describing the sub-barrier resonant dynamics critical to light stellar nucleosynthesis cycles.
\end{abstract}

\maketitle

\section{Introduction}\label{sec1}

The precise determination of low-energy nuclear resonance parameters— specifically resonance energies, lifetimes, and cross-section strengths—is essential for modern nuclear astrophysics \cite{Carlson2017, Mukhamedzhanov2023}. Crucially, these resonances, situated deep within the Gamow window, govern the rates of thermonuclear reactions in stellar environments such as the Sun, asymptotic giant branch (AGB) stars, and classical novae \cite{Newton2007}. In the regime of light nuclei, these states frequently exhibit a delicate interplay between bound-state structure and continuum coupling, presenting a formidable theoretical challenge for few-body nuclear physics \cite{Myo2020, Johnson2020}. The magnitude of this challenge is further underscored by recent experimental inquiries; in several key reactions, newly measured resonance energies, strengths, and decay properties differ significantly from long-standing evaluations. Specific examples of these discrepancies in radiative proton capture reactions are detailed below.

At the lightest end of the mass spectrum, the ${}^7\mathrm{Li}(p,\gamma)^8\mathrm{Be}$ reaction serves as a paradigmatic example, where the low-energy cross-section is dominated by the well-known $1^+$ resonance at $E_p = 441$ keV, corresponding to the $E_x = 17.64$ MeV state in ${}^8\mathrm{Be}$ \cite{Zahnow1995}. Observations of anomalous angular correlations in the internal pair creation ($e^+e-$) decay of this state were initially interpreted as evidence for the existence of a new gauge boson, the X17 particle \cite{Krasznahorkay2016}. However, subsequent theoretical analysis \cite{Hayes2021} argued that the ratio of electromagnetic multipoles—assumed constant across the resonance in the original experiment—actually varies rapidly with energy; accounting for this dependence allows the data to be reconciled with the Standard Model, rendering the X17 hypothesis unnecessary. Conversely, the most recent experimental investigations focusing on the decay of the Giant Dipole Resonance in ${}^8\mathrm{Be}$ reinforce the interpretation of the anomaly as a signature of the X17 particle \cite{Krasznahorkay2024}. Furthermore, similar anomalies have now been reported in other light nuclei, including ${}^4\mathrm{He}$ and $^{12}\mathrm{C}$, suggesting a persistent phenomenon beyond the specific structure of Beryllium \cite{Krasznahorkay2021, Krasznahorkay2022}.

In the CNO mass region, recent findings for reactions such as ${}^{12}\mathrm{C}(p,\gamma)^{13}\mathrm{N}$ and ${}^{14}\mathrm{N}(p,\gamma)^{15}\mathrm{O}$ have similarly superseded legacy data, leading to revised astrophysical scenarios. For instance, in the ${}^{12}\mathrm{C}(p,\gamma)^{13}\mathrm{N}$ reaction, precise new measurements have constrained the width of the prominent 460 keV resonance to $\Gamma=38.2\pm0.5$ keV, thereby firmly establishing the reaction rate for initial carbon burning \cite{Csedreki2023}. In the case of ${}^{14}\mathrm{N}(p,\gamma)^{15}\mathrm{O}$, a re-evaluation of the $E_p = 1058$ keV resonance strength yielded a value 20$\%$ higher than previous recommendations; this shift directly propagates to the solar energy regime, refining predictions for CNO neutrino fluxes \cite{Gyurky2019}.

Turning to reactions involving relatively heavier nuclei, we observe that new experimental data have reached a level of precision that directly impacts stellar nucleosynthesis predictions. A prominent example is the ${}^{21}\mathrm{Ne}(p,\gamma)^{22}\mathrm{Na}$ reaction, a central driver of the NeNa cycle in classical novae, where recent high-precision measurements at LUNA reported resonance strengths for the $272.3$~keV and $352.6$~keV states that exceed previously adopted values by approximately $50\%$, thereby necessitating a recalibration of nova yields \cite{Sidhu2025, Shamsuzzoha2015}. 
Parallel advances have been made for the ${}^{25}\mathrm{Mg}(p,\gamma)^{26}\mathrm{Al}$ reaction, where improved low-energy precision has tightly constrained the production of the cosmic $\gamma$-ray emitter ${}^{26}\mathrm{Al}$ by suppressing cosmic-ray-induced backgrounds \cite{Liu2025}. 
Crucially, the ${}^{23}\mathrm{Na}(p,\gamma)^{24}\mathrm{Mg}$ reaction links these cycles via a sharp resonance at $E_p = 2078$~keV. Recent high-precision measurements have resolved earlier cross-section discrepancies in this channel, providing a robust dataset essential for modeling the NeNa-MgAl transition in AGB stars \cite{Chiar2019}.

As we can see from the above reactions, the interplay between evolving experimental data and theoretical frameworks is particularly evident in the domain of low-energy nuclear resonances. From a fundamental theoretical perspective, these resonances are rigorously defined as Gamow-Siegert states \cite{Michel2002, Michel2003, Dong2022}. In contrast to bound states, the wave functions associated with these S-matrix poles---located in the fourth quadrant of the complex energy plane---exhibit exponentially diverging asymptotic behavior \cite{Mukhamedzhanov2023, Morikawa2025}.  This divergence renders the wave functions non-square-integrable within the standard Hilbert space, causing conventional normalization procedures to fail and necessitating the adoption of Rigged Hilbert Space formulations \cite{Bollini1998, Civitarese2004, Marcucci2016}. Within this framework, physical observables such as the Asymptotic Normalization Coefficient (ANC) can be extracted using specific regularization techniques, most notably Zel'dovich regularization (utilizing a Gaussian convergence factor $\lim_{\epsilon \to 0} e^{-\epsilon r^2}$) \cite{Perelomov1998, Mukhamedzhanov2019, Mukhamedzhanov2022}.

The amplitude of the radiative tail, quantified by the ANC, is analytically connected to the observable resonance width, $\Gamma$. For narrow resonances, such as those in the CNO cycle, this relationship is precise and model-independent, linking the scattering phase shift derivative to the residue of the S-matrix pole \cite{Mukhamedzhanov2010}. However, standard phenomenological approaches like the R-matrix method circumvent the explicit treatment of this continuum coupling by introducing an artificial channel radius, $R_{ch}$, which partitions the space into internal and external regions \cite{Descouvemont2010, Brune2020}. While effective for fitting data, the R-matrix parameters (formal widths and boundary conditions) are often dependent on the arbitrary choice of $R_{ch}$, obscuring the microscopic origin of the resonance formation.

Therefore, a robust theoretical framework must satisfy two stringent criteria: it must treat the resonance mechanism non-perturbatively to accommodate the strong nuclear attraction, while simultaneously preserving the exact analytical structure of the continuum propagator to correctly describe tunneling widths and asymptotic behavior. The real-space Green's function formalism developed herein addresses this dual requirement. By solving the Dyson equation for a delta-shell impurity in a Coulomb field, we circumvent the artificial spatial partitioning of R-matrix theory and the basis-truncation limitations of \textit{ab initio} expansions. Instead, we directly access the spectral density of the system, where resonances emerge naturally as poles of the full propagator, enabling a unified description of resonant states within a single analytical model.

We apply this framework to describe the sub-barrier resonant dynamics governing the proton capture channels for the ${}^7\text{Li}$, ${}^{14}\text{N}$, and ${}^{23}\text{Na}$ systems. We demonstrate that the diagrammatic summation naturally recovers the physical "threshold" and "saturated" regimes of resonance formation, yielding proton resonance energies that exhibit excellent agreement with experimental data as a function of the delta-shell potential strength. Furthermore, our calculations establish the applicability of this formalism for determining resonances in proton capture reactions for nuclei up to $Z \leq 18$.

The present manuscript is organized as follows. In Sec.~\ref{sec2}, we introduce the real-space diagrammatic formalism that constitutes the theoretical foundation of this study. Utilizing bare Green's functions as continuum propagators, we demonstrate how the summation of all possible quantum paths to infinite order recovers the exact transition and reflection coefficients for tunneling through a pure barrier. Furthermore, this section details the calibration of the effective barrier width via the Coulomb interaction, the modeling of the short-range attractive strong nuclear force as an impurity potential, and the derivation of the resonance condition defined by the poles of the full Green's function. The results are presented in Sec.~\ref{sec3}, where we calculate resonance energies, relative cross-sections, and lifetimes for proton capture reactions on ${}^7\text{Li}$, ${}^{14}\text{N}$, and ${}^{23}\text{Na}$ nuclei. Finally, conclusions are provided in Sec.~\ref{sec:conc}.

\section{Real-space diagrammatic description of low-energy nuclear resonances}\label{sec2}

In this work, we establish a theoretical framework starting with the definition of a background field that describes the proton-nucleus interaction—dominated by Coulomb repulsion—prior to the inclusion of the strong nuclear force. To this end, we first formulate a real-space diagrammatic method for tunneling through a pure barrier, effectively mapping the tunneling problem onto a scattering problem involving transitions between distinct potential regions. 
We begin by considering the motion of a particle with energy $E$ as shown in Fig.~\ref{fig:barrier}a, where the potential barrier $V(x)$ is defined as follows:
\begin{equation}\label{eq:potential}
    V(x) = 
    \begin{cases}
    V_0, & \quad a<x<b,\\
    0, & \quad \text{otherwise}.
    \end{cases}
\end{equation}
Since we are investigating quantum tunneling, we consider the case $E<V_0$. In conventional quantum mechanics, by writing the wave functions separately for each interval where the potential (\ref{eq:potential}) is defined---namely $x<a$, $a<x<b$, and $x>b$---and applying the boundary conditions of continuity of the wave functions and their first derivatives at the boundaries, the solution of the stationary Schrödinger equation yields the following expressions \cite{Landau1977}:
\begin{equation}\label{eq:coefficients}
    |T|^2 = \frac{1}{1+\frac{\left(k^2+\kappa^2\right)^2}{4 k^2 \kappa^2} \sinh^2\left(\kappa L\right)}, \quad |R|^2 = 1 - |T|^2,
\end{equation}
where $|T|^2$ and $|R|^2$ denote the transmission and reflection coefficients, respectively, with $L = b-a$, $k = \sqrt{\left(2m/\hbar^2\right)E}$, and $\kappa = \sqrt{\left(2m/\hbar^2\right)(V_0 - E)}$. 

By mapping the problem depicted in Fig.~\ref{fig:barrier}a onto scattering theory via the Fisher--Lee relation, we can write \cite{Fisher1981}:

\begin{figure}
\centering
\includegraphics[width=0.49\textwidth]{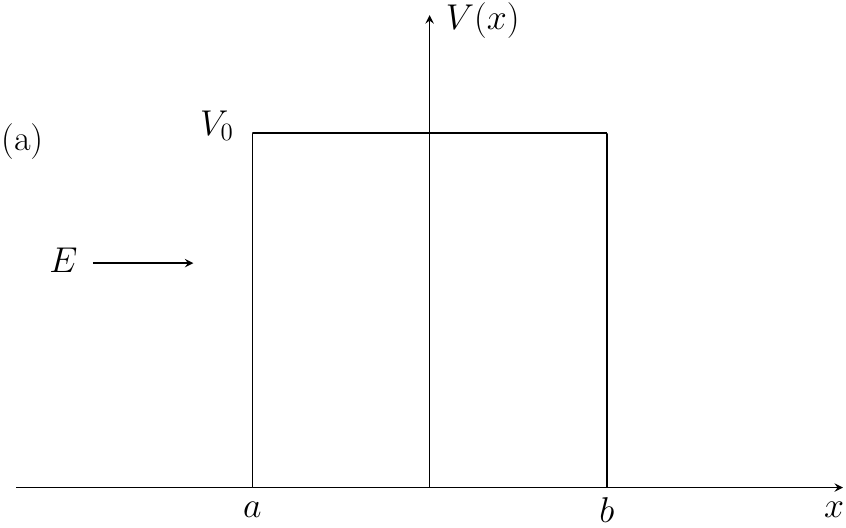}
\hfill
\includegraphics[width=0.49\textwidth]{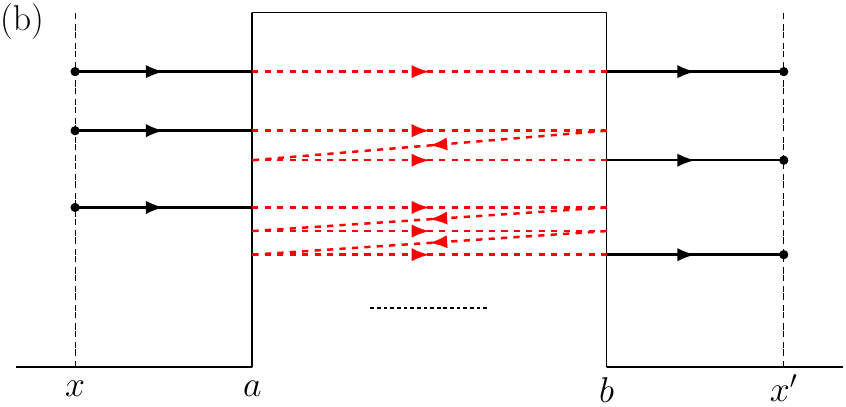}
\hfill
\includegraphics[width=0.49\textwidth]{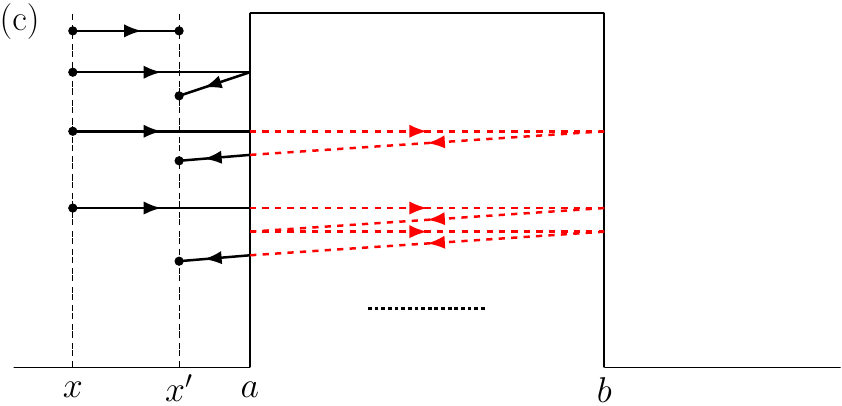}
\hfill
\includegraphics[width=0.49\textwidth]{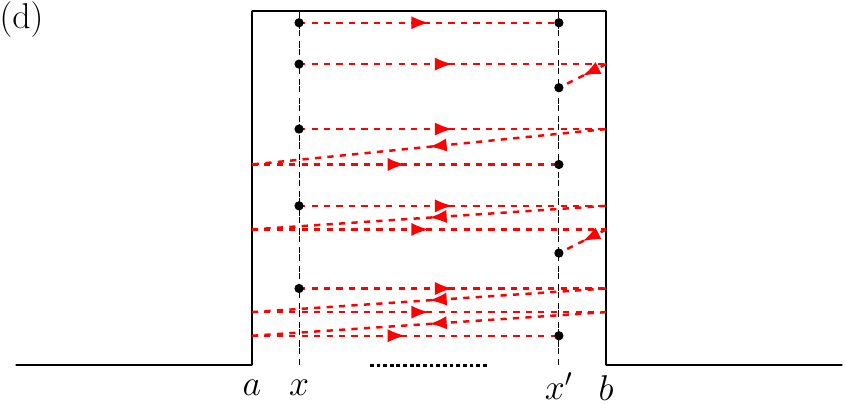}
\caption{Diagrammatic summation of quantum paths for the 1D potential barrier. (a) The potential profile $V(x)$. (b) Paths contributing to the transmission Green's function (Eq.~(\ref{eq:G_tr_1})). (c) Paths contributing to the reflection Green's function (Eq.~(\ref{eq:G_ref_1})). (d) Infinite internal reflections constituting the barrier propagator $\mathcal{D}(x,x'|E)$ (Eq.~(\ref{eq:Green_D})), used in the impurity Dyson equation.}
\label{fig:barrier}
\end{figure}

\begin{equation}\label{eq:Fisher-Lee}
    |T|^2 = \hbar^2 \nu^2 \left|\mathcal{G}_\text{tr} \left(x, x^\prime | E\right)\right|^2, \quad |R|^2 = \hbar^2 \nu^2 \left|\mathcal{G}_\text{ref} \left(x, x^\prime | E\right)\right|^2,  
\end{equation}
where $\nu = \hbar k /m$ is the group velocity of the propagating channel, and $\mathcal{G}_\text{tr} \left(x, x' | E\right)$ and $\mathcal{G}_\text{ref} \left(x, x' | E\right)$ are the retarded Green's functions in the coordinate-energy representation associated with transmission and reflection from the barrier. These Green's functions can be calculated as the result of the summation of infinite-order diagrams shown in Figs.~\ref{fig:barrier}b and \ref{fig:barrier}c. Examining these diagrams, we observe that the entire summation process can be constructed using two types of diagrams: the first type, represented as $\tikz[baseline=-0.5ex]{\path[draw,midarrow,line width=1pt] (0,0)--(0.8,0);}$, corresponds to the bare Green's function associated with the free motion of the particle, while the second type, represented as $\tikz[baseline=-0.5ex]{\path[draw,midarrow,red, dashed, line width=1pt] (0,0)--(0.8,0);}$, corresponds to the bare Green's function associated with the motion of the particle under the potential $V(x) = V_0$. If we denote the first type of bare Green's function as $G_0 \left(x, x'| E \right)$ and the second type as $D_0 \left(x, x'| E \right)$, and label the effective potentials for scattering on the left $(x=a)$ and right $(x=b)$ barrier walls as $V_a$ and $V_b$, respectively, we can write for $\mathcal{G}_\text{tr} \left(x, x' | E\right)$ from Fig.~\ref{fig:barrier}b:
\begin{eqnarray}\label{eq:G_tr_1}
\mathcal{G}_\text{tr} \left(x, x^\prime| E \right)
&=& G_0 \left(x, a| E \right) V_a D_0\left(a,b| E \right) V_b G_0 \left(b, x^\prime| E \right) \nonumber\\
&& + G_0 \left(x, a| E \right) V_a D_0\left(a,b| E \right) V_b D_0\left(b,a| E \right)\nonumber\\
&& \times  V_a D_0\left(a,b| E \right) V_b G_0 \left(b, x^\prime| E \right) \nonumber\\
&& + G_0 \left(x, a| E \right) V_a D_0\left(a,b| E \right) V_b D_0\left(b,a| E \right) \nonumber\\
&& \times V_a D_0\left(a,b| E \right) V_b D_0\left(b,a| E \right) V_a D_0\left(a,b| E \right) \nonumber\\
&& \times V_b G_0 \left(b, x^\prime| E \right) +\dots\nonumber\\
&=& \frac{G_0 \left(x, a| E \right) V_a D_0\left(a,b| E \right) V_b G_0 \left(b, x^\prime| E \right)}{1-D_0\left(b,a| E \right) V_a D_0\left(a,b| E \right) V_b }.\nonumber\\
\end{eqnarray} 
Performing similar calculations for $\mathcal{G}_\text{ref} \left(x, x' | E\right)$ from Fig.~\ref{fig:barrier}c, we find:
\begin{eqnarray}\label{eq:G_ref_1}                
\mathcal{G}_\text{ref} \left(x, x^\prime| E \right) 
&=& G_0 \left(x, x^\prime| E \right) + G_0 \left(x, a| E \right) V_a G_0 \left(a, x^\prime| E \right)  \nonumber\\
&& + G_0 \left(x, a| E \right) V_a D_0 \left(a, b| E \right) V_b D_0 \left(b, a| E \right) \nonumber\\
&& \times  V_a G_0 \left(a, x^\prime| E \right) \nonumber\\
&& + G_0 \left(x, a| E \right) V_a D_0 \left(a, b| E \right) V_b D_0 \left(b, a| E \right) \nonumber\\
&& \times  V_a D_0 \left(a, b| E \right) V_b D_0 \left(b, a| E \right)  \nonumber\\
&& \times V_a G_0 \left(a, x^\prime| E \right) +\dots \nonumber\\
&=& G_0 \left(x, x^\prime| E \right) \nonumber\\
&& + \frac{G_0 \left(x, a| E \right) V_a G_0 \left(a, x^\prime| E \right)}{1-V_a D_0 \left(a, b| E \right) V_b D_0 \left(b, a| E \right)}.
\end{eqnarray}
To find the explicit forms of Eqs.~(\ref{eq:G_tr_1}) and (\ref{eq:G_ref_1}) dependent on coordinate and energy, we first calculate the bare Green's functions for diagrams where the potential (\ref{eq:potential}) is zero. In these regions, the particle is in free motion. Denoting the kinetic energy of the particle as $\hbar^2 {k'}^2/2m$, we can express the bare retarded quantum-mechanical Green's functions in the coordinate-energy representation as follows \cite{Fetter2003}:
\begin{eqnarray}\label{eq:def_G_0}                
G_0 \left(x, x^\prime | E\right) 
&=& \int \frac{dk^\prime}{2 \pi} \frac{e^{i k^\prime \left(x - x^\prime\right)}}{E - \frac{\hbar^2 {k^\prime}^2}{2 m} + i \times 0} \nonumber\\
&=& -  \frac{m}{\pi \hbar^2}  \int dk^\prime \frac{e^{i k^\prime \left(x - x^\prime\right)}}{k^\prime - \sqrt{\left(2m/\hbar^2\right)E} - i \times 0} \nonumber\\
&& \times \frac{1}{k^\prime + \sqrt{\left(2m/\hbar^2\right)E} + i \times 0}.\nonumber\\
\end{eqnarray}
Using the residue theorem in Eq.~(\ref{eq:def_G_0}) for the case $x>x'$, we find \cite{Arfken2013}:
\begin{eqnarray}\label{eq:x>x'}           
G_0 \left(x, x^\prime | E\right) 
&=& -  \frac{m}{\pi \hbar^2} \times 2 \pi i  \frac{e^{ i \sqrt{\left(2m/\hbar^2\right)E} \left(x - x^\prime\right)}}{ 2 \sqrt{\left(2m/\hbar^2\right)E} }\nonumber\\
&\equiv& - \frac{i}{\hbar \nu} e^{ i k \left(x - x^\prime\right)}
\end{eqnarray}
where $\nu = \hbar k / m,$ and $k = \sqrt{\left(2m/\hbar^2\right)E}$. Performing similar calculations for the case $x<x'$, we obtain:
\begin{eqnarray}\label{eq:x<x'}           
G_0 \left(x, x^\prime | E\right) 
&=& -  \frac{m}{\pi \hbar^2} \times (- 2 \pi i)  \frac{e^{- i \sqrt{\left(2m/\hbar^2\right)E} \left(x - x^\prime\right)}}{- 2 \sqrt{\left(2m/\hbar^2\right)E} }\nonumber\\
&\equiv& - \frac{i}{\hbar \nu} e^{- i k \left(x - x^\prime\right)}.
\end{eqnarray}
From Eqs.~(\ref{eq:x>x'}) and (\ref{eq:x<x'}), we find that the bare Green's functions in the case $V(x) = 0$ can be calculated using the following expression:
\begin{equation}\label{eq:G0_main}
    G_0 \left(x, x^\prime | E\right) = - \frac{i}{\hbar \nu} e^{ i k \left|x - x^\prime\right|}.
\end{equation}
Now, let us calculate the bare Green's functions for the case $V(x) = V_0$. Denoting this function as $D_0$, we can write:
\begin{eqnarray}\label{eq:def_D_0}                 
D_0 \left(x, x^\prime | E\right) 
&=& \int \frac{dk^\prime}{2 \pi} \frac{e^{i k^\prime \left(x - x^\prime\right)}}{E - V_0 - \frac{\hbar^2 {k^\prime}^2}{2 m}}  \nonumber\\
&=& -  \frac{m}{\pi \hbar^2}  \int dk^\prime \frac{e^{i k^\prime \left(x - x^\prime\right)}}{k^\prime - i \sqrt{\left(2m/\hbar^2\right)(V_0 - E)} }\nonumber\\
&& \times \frac{1}{k^\prime + i \sqrt{\left(2m/\hbar^2\right)(V_0 - E)}}.
\end{eqnarray}
As seen from Eq.~(\ref{eq:def_D_0}), in the tunneling case we are considering, i.e., $E<V_0$, the addition of $i \times 0$ as in Eq.~(\ref{eq:def_G_0}) is not required when calculating $D_0$. This is because, since $E<V_0$, integration with respect to $k'$ can be performed in the complex plane, as seen in the second line of Eq.~(\ref{eq:def_D_0}). Thus, using the residue theorem, similar to Eq.~(\ref{eq:x>x'}), we can find from Eq.~(\ref{eq:def_D_0}) for the case $x>x'$:
\begin{eqnarray}\label{x>x',D0}                
D_0 \left(x, x^\prime | E\right) 
&=& -  \frac{m}{\pi \hbar^2}  \times 2 \pi i \frac{e^{i \cdot i \sqrt{\left(2m/\hbar^2\right)(V_0 - E)} \left(x - x^\prime\right)}}{2 i \sqrt{\left(2m/\hbar^2\right)(V_0 - E)} } \nonumber\\
&=& -  \frac{1}{\hbar \vartheta}  e^{- \kappa \left(x-x^\prime\right)}, 
\end{eqnarray}
where $\vartheta = \hbar \kappa/m$ and $\kappa = \sqrt{\left(2m/\hbar^2\right)(V_0 - E)}$. For the case $x<x'$, we can write:
\begin{eqnarray}\label{x<x',D0}                
D_0 \left(x, x^\prime | E\right) 
&=& -  \frac{m}{\pi \hbar^2}  \times (- 2 \pi i) \frac{e^{- i \cdot i \sqrt{\left(2m/\hbar^2\right)(V_0 - E)} \left(x - x^\prime\right)}}{- 2 i \sqrt{\left(2m/\hbar^2\right)(V_0 - E)} } \nonumber\\
&=& -  \frac{1}{\hbar \vartheta}  e^{\kappa \left(x-x^\prime\right)}.
\end{eqnarray}
Thus, we obtain the following result for the bare Green's function in the case $V(x) = V_0$:
\begin{equation}\label{eq:D0_main}
    D_0 \left(x, x^\prime | E\right) =  - \frac{1}{\hbar \vartheta}  e^{- \kappa \left|x-x^\prime\right|}.
\end{equation}
Knowing the Green's functions in (\ref{eq:G0_main}) and (\ref{eq:D0_main}), we can calculate the $\mathcal{G}_\text{tr} \left(x, x' | E\right)$ and $\mathcal{G}_\text{ref} \left(x, x' | E\right)$ Green's functions in Eq.~(\ref{eq:Fisher-Lee}). In this case, from Eq.~(\ref{eq:G_tr_1}), we find:
\begin{equation}\label{eq:G_tr_2}
\mathcal{G}_\text{tr} \left(x, x^\prime | E\right) = \frac{\vartheta V_a V_b}{\hbar \nu^2 } \frac{ e^{-ik (b-a+x-x^\prime)}e^{-\kappa L}}{\hbar^2 \vartheta^2- V_a V_b e^{-2 \kappa L}}.
\end{equation}
Similarly, from Eq.~(\ref{eq:G_ref_1}), we find:
\begin{equation}\label{eq:G_ref_2}
\mathcal{G}_\text{ref} \left(x, x^\prime | E\right) = - \frac{i}{\hbar \nu} e^{i k (x^\prime - x)}-\frac{\vartheta^2 V_a}{ \nu^2}
 \frac{e^{i k (2a - x - x^\prime)}}{\hbar^2 \vartheta^2 - V_a V_b e^{ -2\kappa L}}.
\end{equation}
Substituting Eqs.~(\ref{eq:G_tr_2}) and (\ref{eq:G_ref_2}) into Eq.~(\ref{eq:Fisher-Lee}), we obtain for the transmission and reflection coefficients:
\begin{equation}\label{eq:T}
    |T|^2 = \frac{\vartheta^2 }{\nu^2} \frac{\left|V_a\right|^2 \left|V_b\right|^2 e^{-2 \kappa L}}{\left(\hbar^2 \vartheta^2- V_a V_b e^{-2 \kappa L}\right)\left(\hbar^2 \vartheta^2- V_a^* V_b^* e^{-2 \kappa L}\right)},
\end{equation}
and
\begin{equation}\label{eq:R}
     |R|^2 = 1 
    + \frac{\hbar^2 \vartheta^4 }{ \nu^2}  \frac{\left|V_a\right|^2 + i \hbar \nu \left[V_a^* - V_a +  \frac{\left|V_a\right|^2}{\hbar^2 \vartheta^2} \left(V_b^* - V_b \right) e^{-2 \kappa L} \right]}{\left(\hbar^2 \vartheta^2- V_a V_b e^{-2 \kappa L}\right)\left(\hbar^2 \vartheta^2- V_a^* V_b^* e^{-2 \kappa L}\right)}.
\end{equation}
Using Eqs.~(\ref{eq:T}) and (\ref{eq:R}), we will derive the equations of conventional quantum mechanics given in Eq.~(\ref{eq:coefficients}). For this, appropriate expressions for the effective potentials $V_a$ and $V_b$ must be determined. Finding these will also indicate that the real-space diagrammatic technique we have developed is working correctly. To this end, let us first consider Eqs.~(\ref{eq:T}) and (\ref{eq:R}) alongside the conventional quantum mechanical equation $|T|^2 + |R|^2 = 1$. From these equations, we can find:
\begin{eqnarray}\label{eq:mon}
    \left(|V_a|^2 + i \hbar \nu \left(V_a^* - V_a\right) \right) \hbar^2 \vartheta^2  - \left(|V_b|^2 + i \hbar \nu \left(V_b - V_b^*\right) \right)\nonumber\\
    \times |V_a|^2 e^{-2\kappa L} = 0.
\end{eqnarray}
If we assume $V_a = V_b^*$ in Eq.~(\ref{eq:mon}), since $\hbar^2 \vartheta^2 - |V_b|^2 e^{-2\kappa L} = 0$ is not allowed by Eqs.~(\ref{eq:T}) and (\ref{eq:R}), we find that $|V_b|^2 + i \hbar \nu \left(V_b - V_b^*\right) = 0$. If we write $V_b = {V_b}_1 + i {V_b}_2$ in this expression, we obtain ${V_b}_1 = \sqrt{2 \hbar \nu {V_b}_2 - {V_b}_2^2}$. Thus, when $V_a = V_b^*$, we can find from Eq.~(\ref{eq:T}):
\begin{equation}\label{eq:mon2}
    |T|^2 = \left(\frac{2k\kappa {V_b}_2}{\frac{\hbar^2}{m} k \kappa^2e^{\kappa L} - 2 k^2 {V_b}_2 e^{-\kappa L}}\right)^2.
\end{equation}
Comparing Eq.~(\ref{eq:mon2}) with the expression for $|T|^2$ from Eq.~(\ref{eq:coefficients}), we find:
\begin{equation}\label{eq:main}
    {V_b}_2 = \frac{\hbar^2}{m}\frac{k \kappa^2 e^{2\kappa L}}{2k^2 +\sqrt{4k^2\kappa^2 + \left(k^2 + \kappa^2\right)^2 \sinh^2(\kappa L)}}.
\end{equation}
Thus, by finding the equations for the effective potentials, we have demonstrated that our diagrammatic technique works correctly. It should be noted that this part of our work is not limited to nuclear physics but is generally applicable to any quantum system involving tunneling.

Since the tunneling problem for a pure barrier has been exactly mapped onto scattering theory, we can directly apply these results to the sub-barrier proton-nucleus scattering system.
To ensure our 1D square barrier model accurately reproduces the tunneling physics of the realistic 3D Coulomb potential, we calibrate the barrier parameters using the WKB approximation. This calibration is justified because the WKB method reproduces the tunneling exponent with high accuracy for smooth potentials like the Coulomb barrier, allowing us to retain the algebraic simplicity of the square barrier formalism without sacrificing physical realism. The detailed formulation of this procedure is provided below.

The sub-barrier tunneling probability is dominated by the exponential Gamow factor, $T \approx e^{-2\mathcal{S}}$, where the action $\mathcal{S}$ is defined by the integral of the momentum through the classically forbidden region.
For a realistic proton-nucleus system, the Coulomb barrier extends from the nuclear radius $x_1$ to the classical outer turning point $x_2$, defined by $E = Z_1 Z_2 e^2 / x_2$. The WKB action for this Coulomb potential is given by:
\begin{eqnarray}
    \mathcal{S}_{\text{Coulomb}} &=& \int_{x_1}^{x_2} \sqrt{\frac{2m}{\hbar^2}\left( \frac{Z_1 Z_2 e^2}{x} - E\right)} dx \nonumber\\
    &=& x_2 k \left[ \frac{\pi}{2} - \sin^{-1}\sqrt{\frac{x_1}{x_2}} - \sqrt{\frac{x_1}{x_2}\left(1-\frac{x_1}{x_2}\right)} \right],\nonumber\\
\label{eq:gamma_coulomb}
\end{eqnarray}
where $k = \sqrt{2mE}/\hbar$. On the other hand, for the square barrier model with height $V_0$ and width $L = b-a$, the WKB action is $\mathcal{S}_{\text{Square}} = L \kappa$, where $\kappa = \sqrt{2m(V_0 - E)}/\hbar$. By equating the tunneling probabilities (Gamow factors) of the two models, $\mathcal{S}_{\text{Square}} = \mathcal{S}_{\text{Coulomb}}$, we derive an energy-dependent effective width $L(E)$:
\begin{equation}
    L(E) = \frac{x_2 k}{\kappa} \left[ \frac{\pi}{2} - \sin^{-1}\sqrt{\frac{x_1}{x_2}} - \sqrt{\frac{x_1}{x_2}\left(1-\frac{x_1}{x_2}\right)} \right].
\label{eq:L_eff}
\end{equation}

Finally, to capture the formation of nuclear resonance states within this framework, we introduce a short-range attractive potential $V_\gamma(x)$, localized at the nuclear radius $x_1$, which corresponds to the interface $b$ in our barrier model. We model this interaction mathematically as a delta-shell potential:
\begin{equation}\label{eq:impurity}
V_\gamma(x) = \gamma \delta(x - x_1),
\end{equation}
where $\gamma$ (in $\text{MeV}\cdot\text{fm}$) represents the integrated strength of the nuclear mean field acting at the surface. For attractive interactions leading to resonance formation, we take $\gamma < 0$. Treating this delta-shell potential as an impurity and denoting the full Green's function of the combined system (Barrier + Impurity) as $\mathbb{G}$, we can invoke the Dyson equation to write \cite{Fetter2003}:
\begin{equation}\label{eq:Green_res_1}
\mathbb{G}(x_1, x_1; E) = \frac{\mathcal{D}(x_1, x_1; E)}{1 - \gamma \mathcal{D}(x_1, x_1; E)},
\end{equation}
where the Green's function $\mathcal{D}$ is calculated as the infinite summation of the diagrams depicted in Fig.~\ref{fig:barrier}d, as follows:
\begin{eqnarray}\label{eq:Green_D}
    \mathcal{D} (x,x^\prime | E) &=& D_0 \left(x, x^\prime | E\right) + D_0 \left(x, b | E\right) V_b D_0 \left(b, x^\prime | E\right) \nonumber\\
    &&+ D_0 \left(x, b | E\right) V_b D_0 \left(b, a | E\right) V_a D_0 \left(a, x^\prime | E\right) \nonumber\\
    && + D_0 \left(x, b | E\right) V_b D_0 \left(b, a | E\right) V_a D_0 \left(a, b | E\right) \nonumber\\
    && \times V_b D_0 \left(b, x^\prime | E\right) \nonumber\\
    &&+ D_0 \left(x, b | E\right) V_b D_0 \left(b, a | E\right) V_a D_0 \left(a, b | E\right) \nonumber\\
    && \times V_b D_0 \left(b, a | E\right) V_a D_0 \left(a, x^\prime | E\right) + \dots\nonumber\\
    &=& D_0 \left(x, x^\prime | E\right) + D_0 \left(x, b | E\right) V_b\nonumber\\
    && \times \frac{D_0 \left(b, x^\prime | E\right) + D_0 \left(b, a | E\right) V_a D_0 \left(a, x^\prime | E\right)}{1-V_a D_0 \left(a, b| E \right) V_b D_0 \left(b, a| E \right)}.\nonumber\\
\end{eqnarray}
Substituting Eq.~(\ref{eq:D0_main}) into the expression yields: 
\begin{equation}\label{eq:Green_inside}
    \mathcal{D} (x,x^\prime | E) = -\frac{1}{\hbar \vartheta} e^{\kappa (x-x^\prime)} + \frac{V_b e^{-\kappa (2b - x - x^\prime) + \frac{V_a V_b}{\hbar \vartheta} e^{-\kappa (2L - x - x^\prime) }}}{\hbar^2 \vartheta^2 - V_a V_b e^{-2 \kappa L}}.
\end{equation}
Thus, having determined $\mathcal{D}$ via Eq.~(\ref{eq:Green_inside}), we can extract the complete physics of the resonance from the denominator of Eq.~(\ref{eq:Green_res_1}) using the following condition:
\begin{equation}\label{eq:res_condition}
1 - \gamma \mathcal{D}(x_1, x_1; \mathcal{E}_{res}) = 0.
\end{equation}
The complex root is decomposed as $\mathcal{E}_{res} = E_{res} - i \Gamma / 2$, where $E_{res}$ is the physical resonance energy and $\Gamma$ is the resonance width. From these values, the lifetime $\tau$ of the resonance state is calculated via the uncertainty principle:
\begin{equation}\label{eq:tau}
\tau = \frac{\hbar}{\Gamma}.
\end{equation}

\section{Results}\label{sec3}

Having established the exact analytical form of the background Green's functions for the potential barrier, we now proceed to apply this formalism to physical nuclear systems. 
The power of the real-space diagrammatic method lies in its ability to treat the strong nuclear interaction non-perturbatively. We model the physical system as a superposition of a background field (Coulomb barrier plus centrifugal potential) and a short-range strong interaction. By introducing a localized "impurity" potential (\ref{eq:impurity}) to represent the strong force at the nuclear surface, we can utilize the Dyson equation (\ref{eq:Green_res_1}) to solve for the full propagator of the interacting system. This allows us to calculate critical physical observables—specifically resonance energies, scattering cross-sections, and lifetimes—by analyzing the pole structure and spectral density of the full Green's function.
To investigate the formation and stability of these resonant states, we applied our formalism to three distinct light nuclear systems: $^7$Li, $^{14}$N and $^{23}$Na. These isotopes were specifically selected to span a representative range of atomic numbers ($Z=3,7,$ and $11$), allowing us to validate the formalism against increasing Coulomb barrier heights while addressing critical resonances that govern the solar p-p chain, the CNO cycle, and the transition to the NeNa-MgAl network.

\subsection{Resonance energy}

To calculate the resonance energies for the $p+{}^7\text{Li}$, $p+{}^{14}\text{N}$, and $p+{}^{23}\text{Na}$ systems by solving Eq.~(\ref{eq:res_condition}), we adopt a standard nuclear radius parameter of $r_0 = 1.2$~fm to define the interaction geometry. Using the classical expression for the Coulomb barrier height, $V_C = Z_p Z_\text{target} e^2 / (r_0 A^{1/3})$, this parametrization yields barrier heights of approximately $1.9$~MeV for ${}^7\text{Li}$, $3.5$~MeV for ${}^{14}\text{N}$, and $4.6$~MeV for ${}^{23}\text{Na}$.

\begin{figure}
    \centering
    \includegraphics[width=\linewidth]{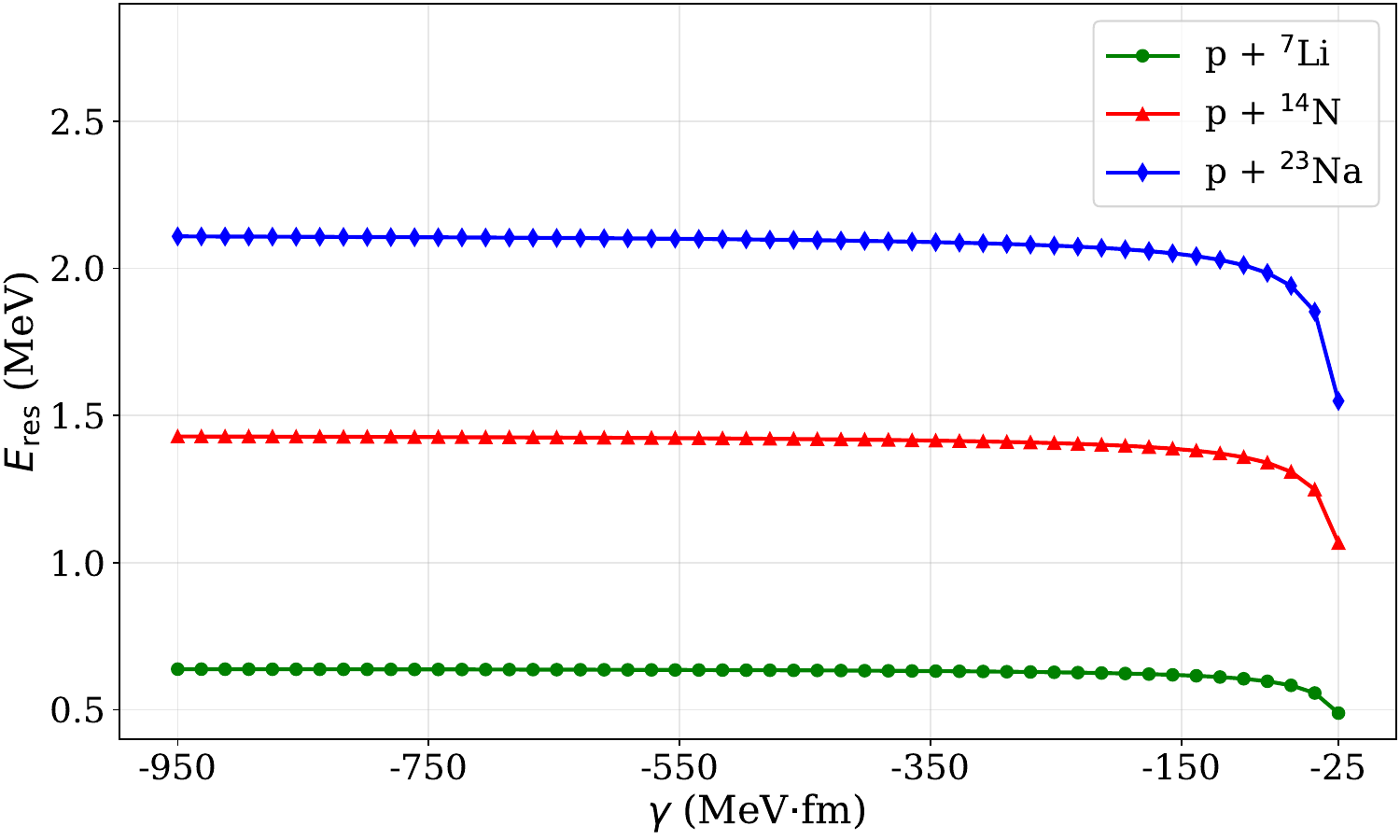} 
    \caption{The dependence of the proton resonance energy $E_{\text{res}}$ on the surface impurity strength $\gamma$ for $^7$Li, $^{14}$N, and $^{23}$Na.}    
    \label{fig:gamma_scan}
\end{figure}

Fig.~\ref{fig:gamma_scan} illustrates the trajectory of the resonance energy as a function of $\gamma$. In the strong-coupling domain ($-950 < \gamma < -350$ MeV$\cdot$fm), we observe a "plateau" where the resonance energies saturate at values determined largely by the barrier geometry. In this saturation regime, the energies stabilize at approximately $0.64$ MeV for Li, $1.43$ MeV for N, and $2.11$ MeV for Na.

Comparison with experimental data reveals a notable physical distinction between the lighter and heavier systems. Specifically, for the $p+{}^7\text{Li}$ and $p+{}^{14}\text{N}$ cases, the saturation plateau overbinds the system relative to the physical states. Theoretical predictions converge to experiment only in the weak-coupling window ($\gamma \approx -25$ MeV$\cdot$fm). At this coupling strength, our model yields $0.489$ MeV for Li (Exp: $0.441$ MeV \cite{Hayes2021}) and $1.067$ MeV for N (Exp: $1.058$ MeV \cite{Gyurky2019}). This suggests that these resonances are "threshold states", sustained by a relatively weak surface interaction that is just sufficient to create a metastable pocket. In contrast, for the $p+{}^{23}\text{Na}$ system, the resonance formation does not require fine-tuning of the coupling strength. The experimental value of $2.08$ MeV \cite{Chiar2019} aligns almost perfectly with our calculated saturation plateau value of $2.11$ MeV (valid for all $|\gamma| > 350$).

This dichotomy highlights a fundamental transition in the physics of resonance formation. For the heavier sodium system, the deeper nuclear well and higher Coulomb barrier ($4.6$ MeV) create a robust geometric environment where the resonance energy is determined primarily by the confinement length $L_{\text{eff}}$ rather than the fine details of the surface interaction strength. The state is effectively "saturated" by the barrier geometry. Conversely, the lighter systems ($^7$Li, $^{14}$N) are more sensitive to the interaction details, residing on the "slope" of the stability curve where the resonance energy is strongly suppressed by the proximity to the continuum threshold.

\subsection{Relative Resonant Cross-Section}

Having identified the distinct coupling regimes where theoretical resonance energies align with experiment---the weak-coupling domain ($\gamma \approx -25$ MeV$\cdot$fm) for the threshold states of $^7$Li and $^{14}$N, and the strong-coupling saturation limit ($\gamma < -350$ MeV$\cdot$fm) for $^{23}$Na---we now examine the spectral signatures of these states. While the resonance energy $E_{\text{res}}$ defines the static stability, the dynamical scattering observables are described in the relative cross-section, $\sigma(E)$. This quantity is defined via the inverse Dyson determinant:
\begin{equation}\label{eq:cs}
    \sigma(E) = \frac{1}{|1 - \gamma \mathcal{D} (E)|^2}.
\end{equation}
Physically, $\sigma(E)$ represents a dimensionless enhancement factor, quantifying the ratio of the resonant scattering probability to the non-resonant background. In this dimensionless parameter, peaks correspond to the energies where the tunneling probability is maximally enhanced by the nuclear attraction.

\begin{figure}
    \centering
    \includegraphics[width=\textwidth]{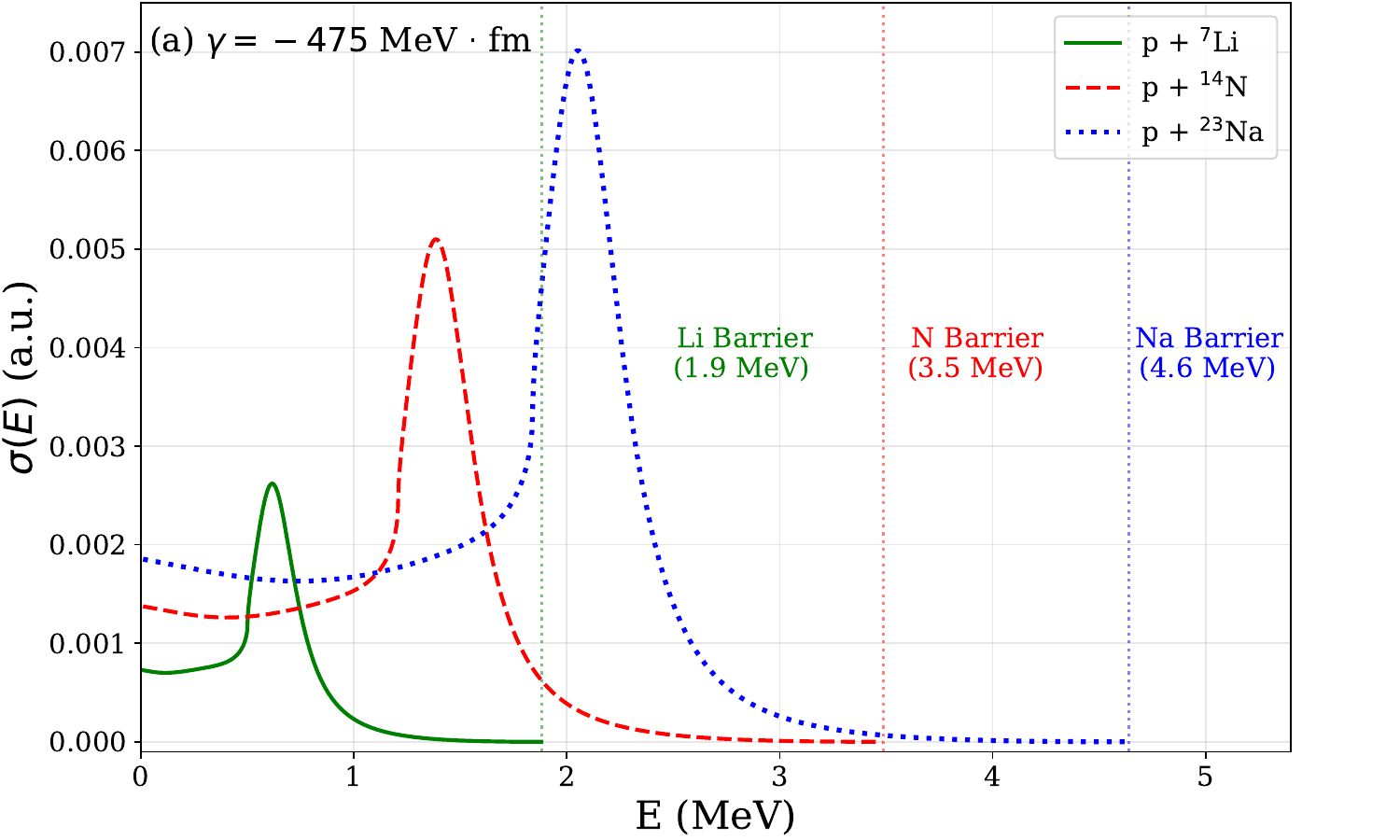}
    \hfill
    \includegraphics[width=\textwidth]{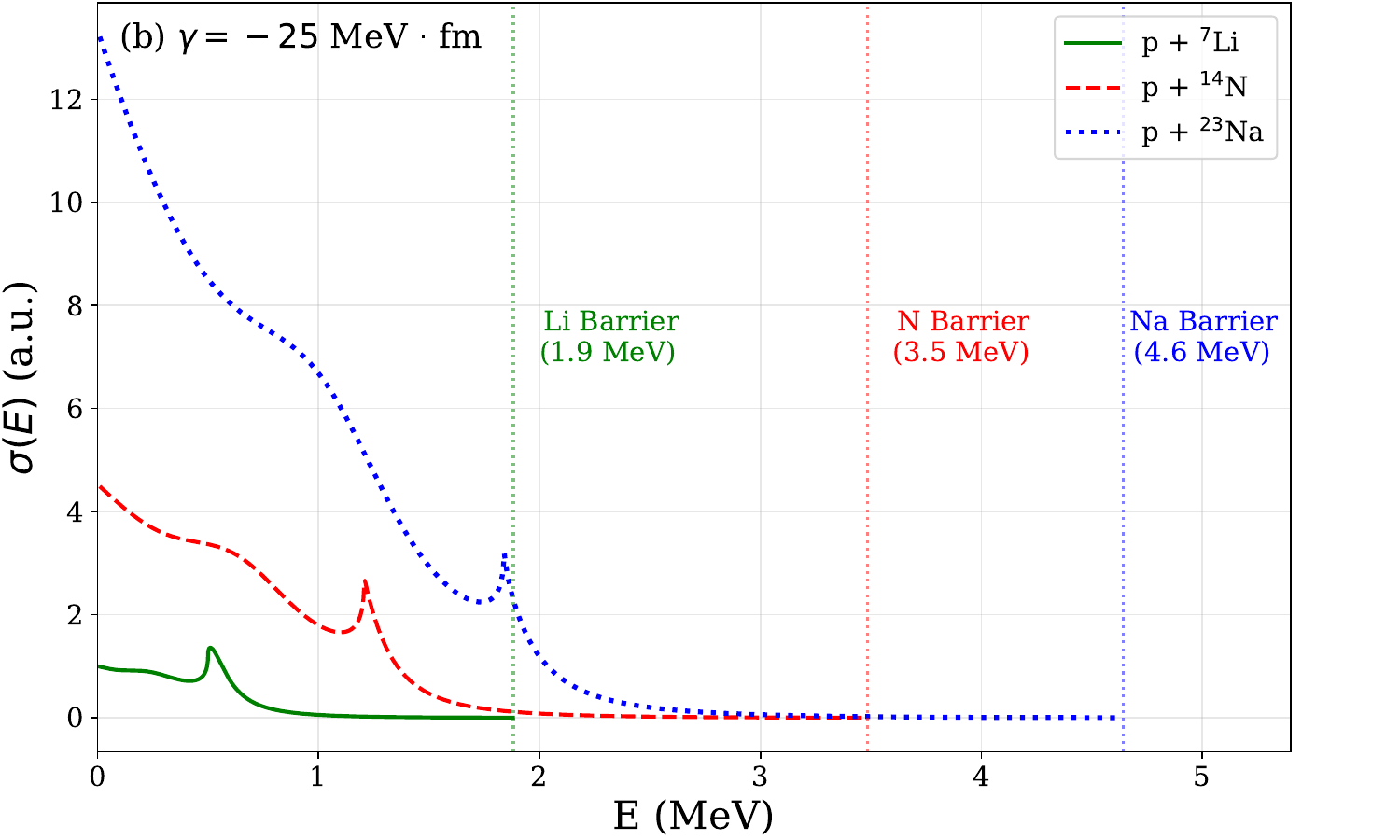}
    \caption{The relative resonant cross-section $\sigma(E)$ for $p+^7$Li (blue), $p+^{14}$N (red), and $p+^{23}$Na (green). (a) In the strong-coupling regime, resonances occur at the saturation "plateau" energies. This regime correctly reproduces the physical $^{23}$Na resonance at $\approx 2.1$ MeV, though with a relatively broad, low-amplitude signature. (b) In the weak-coupling regime, the Li and N resonances shift to their physical energies ($0.49$ MeV and $1.07$ MeV) and exhibit a massive enhancement in amplitude, characteristic of sharp threshold states. The vertical dotted lines indicate the classical Coulomb barrier heights.}
    \label{fig:cross_section}
\end{figure}

Fig.~\ref{fig:cross_section} presents the calculated relative cross-section profiles for $p + {}^7\text{Li}$, $p + {}^{14}\text{N}$, and $p + {}^{23}\text{Na}$ in the two limiting interaction regimes: the strong-coupling saturation limit ($\gamma = -475$ MeV$\cdot$fm, panel a) and the physical weak-coupling limit for light nuclei ($\gamma = -25$ MeV$\cdot$fm, panel b).
The results demonstrate a marked difference in scattering magnitude between the saturated and threshold states, a behavior fully predicted by the analytic structure of our theory. Specifically, in the strong-coupling case (Fig.~\ref{fig:cross_section}a), all systems appear at their geometric saturation energies. Crucially, this represents the physical regime for $p+^{23}$Na, which exhibits a distinct peak at $\approx 2.1$ MeV, matching the experimental resonance \cite{Chiar2019}. However, the peak amplitude remains relatively suppressed (e.g., $\sigma \sim 0.007$ a.u. for $p+{}^{23}\text{Na}$). In contrast, the weak-coupling case (Fig.~\ref{fig:cross_section}b) corresponds to the physical regime for $p + {}^7\text{Li}$ and $p + {}^{14}\text{N}$, where the peaks not only shift to the correct experimental energies, but their amplitudes also exhibit a notable enhancement, with peak values generally exceeding $\sigma > 1.0$ a.u.

This magnitude disparity is physically justified by the analytic properties of the resonant amplitude. At resonance, the peak cross-section is inversely proportional to the square of the imaginary part of the propagator:
\begin{equation}
    \sigma(E_{\text{res}}) \approx \frac{1}{|\gamma \text{Im}[\mathcal{D}(E_{\text{res}})]|^2}.
\end{equation}
In the weak-coupling regime characteristic of the lithium and nitrogen states, two factors constructively interfere to maximize this signal. First, the reduction in the coupling strength $\gamma$ directly reduces the denominator, amplifying the ratio. Second, as these resonances shift to lower energies near the threshold, the tunneling probability—represented by $\text{Im}[\mathcal{D}]$—decreases exponentially due to the thickening of the Coulomb barrier. Since this term appears in the denominator, the suppression of the tunneling width results in a substantial enhancement of the resonant peak height. This mechanism renders the threshold states significantly sharper and more pronounced than their saturated counterparts, which reside deeper in the continuum where tunneling is less suppressed.

\subsection{Domain of Validity}

While the resonance energy $E_{\text{res}}$ locates the spectral position and the relative cross-section $\sigma(E)$ describes the scattering intensity, the dynamical stability of the system is best quantified by its lifetime, $\tau$. In the context of scattering, this corresponds to the time delay experienced by the proton while it is temporarily trapped inside the nuclear potential barrier. In our diagrammatic framework, this observable is extracted directly from the imaginary part of the resonance pole via the uncertainty relation~(\ref{eq:tau}). This metric serves as a critical diagnostic for establishing the domain of validity for our scattering formalism.

\begin{figure}
    \centering
    \includegraphics[width=\linewidth]{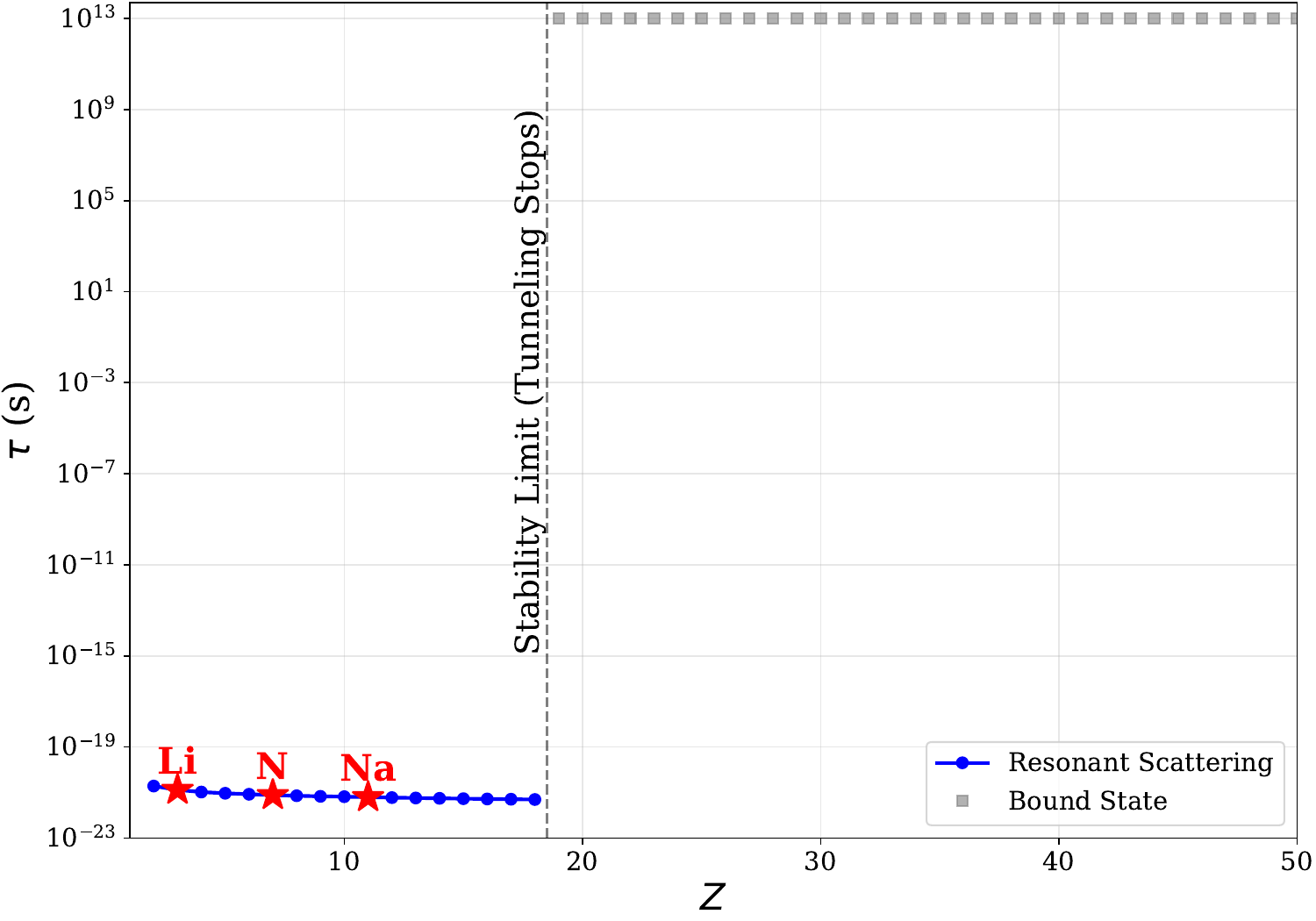} 
    \caption{The transition from quantum scattering to classical stability. For light nuclei ($Z \le 18$), the $p+\text{Nucleus}$ system supports broad scattering resonances with lifetimes on the order of $10^{-21}$~s (blue circles). A distinct stability limit is reached at Argon ($Z = 18$). Beyond this vertical cutoff, the Gamow suppression becomes severe, causing the lifetime to diverge to macroscopic scales. This indicates that for $Z > 18$, the proton is effectively trapped, forming a bound state rather than a transient scattering resonance.}
    \label{fig:lifetime_transition}
\end{figure}

Fig.~\ref{fig:lifetime_transition} maps the calculated resonance lifetime across the chart of light-to-medium nuclides ($Z=2 \to 50$) using an interaction strength of $\gamma = -475$~MeV$\cdot$fm to effectively illustrate the dynamical range of the formalism. This metric serves as a critical diagnostic for establishing the domain of validity for the scattering model. In this context, we observe from Fig.~\ref{fig:lifetime_transition} that for light nuclei, ranging from Helium ($Z=2$) to Argon ($Z=18$), the system resides firmly in the quantum tunneling regime. As shown in the lower-left quadrant of Fig.~\ref{fig:lifetime_transition} (blue trajectory), the calculated lifetimes cluster in the range of $\tau \sim 10^{-22} - 10^{-21}$~s. This timescale represents the duration of the intermediate compound state formed during the collision (e.g., the formation of a short-lived ${}^8\text{Be}^*$ state in the $p+{}^7\text{Li}$ reaction). Although these calculated values represent the idealized single-particle limit, they fall within the same order of magnitude as experimental observations, confirming that the model correctly describes the temporal scale of the tunneling process. Notably, the specific benchmarks for astrophysical cycles fall squarely in this window:
\begin{itemize}
    \item $p + {}^{7}\text{Li} \, (Z=3)$: $\tau \approx 1.30 \times 10^{-21}$~s
    \item $p + {}^{14}\text{N} \, (Z=7)$: $\tau \approx 7.69 \times 10^{-22}$~s
    \item $p + {}^{23}\text{Na} \, (Z=11)$: $\tau \approx 6.15 \times 10^{-22}$~s
\end{itemize}
The relatively flat trajectory of the lifetime in this region indicates that the resonant states remain broad and physically accessible, confirming that the diagrammatic summation method is ideally suited for describing the reaction dynamics of light stellar cycles. In contrast, a sharp physical discontinuity is observed at $Z \approx 18$ (dashed vertical line). As the atomic number increases, the height and width of the Coulomb barrier grow non-linearly, leading to an exponential suppression of the tunneling probability. At the critical threshold of Argon, the resonance width $\Gamma$ drops below the scale of physical observability.

For nuclei heavier than Argon ($Z > 18$), the calculated lifetimes diverge to macroscopic scales, indicated by the gray squares in the upper region of Fig.~\ref{fig:lifetime_transition}. In this regime, the interaction is no longer scattering in the resonant sense; if the proton were to enter the nucleus, the barrier would prevent it from tunneling back out. Thus, the system forms a quasi-stable or bound state. Consequently, we identify $Z = 18$ as the upper bound for the applicability of this specific real-space scattering method.

\section{Conclusion}\label{sec:conc}

In this work, we have developed and applied a non-perturbative, real-space diagrammatic framework to describe proton-nucleus resonance scattering. By mapping the quantum tunneling problem onto a scattering formalism involving the summation of infinite diagrammatic series, we derived exact analytical expressions for the Green's functions of a system governed by a Coulomb barrier and a localized strong interaction. This approach allowed us to treat the resonance formation not as a perturbative correction, but as a pole of the full propagator derived from the Dyson equation.

We validated the formalism by calculating the spectroscopic properties of proton scattering on astrophysically significant nuclei: $p + ^7$Li, $p + ^{14}$N, and $p + ^{23}$Na. Our analysis revealed a fundamental physical dichotomy in resonance formation. For the lighter systems ($^7$Li and $^{14}$N), the physical resonances emerge in a distinct weak-coupling regime ($\gamma \approx -25$ MeV$\cdot$fm). In this domain, theoretical predictions converge with high accuracy to experimental benchmarks ($0.49$ MeV for Li, $1.07$ MeV for N) \cite{Hayes2021, Gyurky2019}. In contrast, the heavier $^{23}$Na system behaves as a saturated state: its experimental resonance at $2.08$ MeV \cite{Chiar2019} aligns with the model's strong-coupling plateau ($2.11$ MeV), indicating that its stability is determined primarily by the barrier geometry rather than the fine-tuning of the surface interaction.

The study of dynamical scattering observables further clarified this distinction. The weak-coupling regime (Li, N) is characterized by a "resonant explosion" of the relative cross-section, where the scattering amplitude increases by orders of magnitude due to the suppression of the tunneling width. Conversely, the strong-coupling regime (Na) exhibits a broader, lower-amplitude spectral signature, consistent with a system where the strong surface impurity effectively hardens the boundary.

Finally, we established the global domain of validity for this method via a systematic scan of resonance lifetimes across the periodic table ($Z=2 \to 50$). In the regime of light nuclei ($Z \le 18$), including Li, N, and Na, the system exhibits characteristic quantum tunneling behavior with observable lifetimes ($\tau \sim 10^{-22}$ s). However, beyond this critical threshold, the exponential Gamow suppression closes the tunneling window, causing lifetimes to diverge to macroscopic scales. This discontinuity marks the physical transition from the quantum scattering regime to the classical limit of stable nuclear isomers. Consequently, while the diagrammatic formalism proves to be a precision tool for the spectroscopy of nuclei relevant to the CNO and NeNa cycles, the description of proton emission in heavier systems ($Z > 18$) requires an asymptotic bound-state approach.

\bibliographystyle{elsarticle-num}
\bibliography{az_bib}
\end{document}